\documentclass[
 reprint,
 amsmath,amssymb,
 aps,
]{revtex4-2}

\usepackage{graphicx}% Include figure files
\usepackage{dcolumn}% Align table columns on decimal point
\usepackage{bm}% bold math
\begin{document}

\preprint{APS/123-QED}

\title{Partially isophase electro-magnetic configurations}% Force line breaks with \\
%\thanks{A footnote to the article title}%

\author{Andrey Pupasov Maksimov}
\affiliation{%
 Universidade Federal de Juiz de Fora\\
 Brasil %\textbackslash\textbackslash
}%

%\collaboration{MUSO Collaboration}%\noaffiliation

%\collaboration{CLEO Collaboration}%\noaffiliation

\date{\today}% It is always \today, today,
             %  but any date may be explicitly specified

\begin{abstract}
Separation of variables in Schr\"{o}dinger equation allows an infinite-dimensional family of physically distinct field configurations that are dynamically indistinguishable in one selected sector, labeled by a conserved quantum number, while differing in all others. We call such a symmetry a partial isophase symmetry. Two main examples are considered: axially symmetric magnetic backgrounds (a rigid spectral flow of the Fock--Darwin oscillator, purely magnetic isophase families, and a two-channel extension) and a Cartesian slab geometry (magnetic barriers in two-dimensional electron gases, treated in the Appendices). We further discuss the consequences of this symmetry for the inverse spectral problem and consider its quasiclassical counterpart.
\end{abstract}

%\keywords{Suggested keywords}%Use showkeys class option if keyword
                              %display desired
\maketitle

%\tableofcontents

\section{\label{sec:Intro} Introduction}

The quantum dynamics of a charged particle in an external electromagnetic field is governed by gauge-invariant combinations of the scalar and vector potentials
\cite{sakurai1994modern}. Standard gauge transformations leave the electromagnetic fields unchanged while modifying the wave-function phase. 
In this Letter, we identify a different type of invariance
which we call a partial isophase symmetry.
This transformation changes the electromagnetic field itself, yet preserves exactly the effective Hamiltonian, and therefore the full quantum evolution, in one selected sector singled out by the symmetries of the problem.

A partial isophase symmetry is a generic consequence of separation of variables
which in turn is related with the symmetry algebra of the differential operators \cite{miller1977symmetry}. 
Whenever continuous symmetries commuting with the Hamiltonian decompose the wave function into sectors 
labeled by conserved quantum numbers, electromagnetic potentials can be varied leaving the corresponding 1D effective potential fixed in exactly one sector, at the price of shifting every other sector. This mechanism is worked out below first for the physically richest example, an axially symmetric background, where the two symmetry directions are a rotation and a translation; Sec.~\ref{sec:general} then abstracts it to an arbitrary separable geometry.
Our findings followed the opposite direction. Having at hand a variety of non-trivial exactly solvable examples of both stationary and non-stationary one-dimensional Schr\"{o}dinger equations, we were interested in more realistic 3D applications for a charged particle in an electromagnetic field. 
However, such an embedding lacks uniqueness. While analyzing the ambiguities that appear, the partial isophase symmetry was discovered. This symmetry is not restricted to exactly solvable models. 

Consider the three-dimensional Schrödinger equation
%\begin{equation}
%(i\hbar \partial_t-eA_0)\Psi(\mathbf r,t)
%=
%\frac{1}{2m}\left(\mathbf p-\frac{e}{c}\mathbf A\right)^2\Psi(\mathbf r,t),
%\label{eq:Sch3D}
%\end{equation}
\begin{equation}
(i\hbar \partial_t-e \phi)\Psi(\mathbf r,t)
=
\frac{1}{2m}\left(\mathbf p-e \mathbf A\right)^2\Psi(\mathbf r,t),
\label{eq:Sch3D}
\end{equation}
in axially symmetric electromagnetic backgrounds that are translation invariant along the $z$ axis \cite{bagrov2012quantum}. 
Such configurations are relevant, in particular, to the dynamics of electron states carrying orbital angular momentum, including vortex electrons, whose controlled generation and propagation has become an active topic in both fundamental and applied research \cite{karlovets2026angular}.

For $z$-independent fields, the potentials can be parametrized in cylindrical coordinates as
%\begin{equation}
%\frac{e}{c\hbar}A_0=f_0(\rho),\qquad
%\frac{e}{c\hbar}\mathbf A=\frac{f_2(\rho)}{\rho}\mathbf e_\phi-f_1(\rho)\mathbf k,
%\label{eq:potparam}
%\end{equation}
% -f_1(\rho)->A_z(\rho)
% f_0(\rho)=\phi(\rho)
%f_2(\rho)=f(\rho)
\begin{equation}
\phi=\phi(\rho),\qquad
\mathbf A=\frac{f(\rho)}{\rho}\mathbf e_\varphi+A_z(\rho)\mathbf e_z,
\label{eq:potparam}
\end{equation}
so that
%\begin{equation}
%\frac{e}{c\hbar}\mathbf E=-\partial_\rho f_0\,\mathbf e_\rho,
%\qquad
%\frac{e}{c\hbar}\mathbf B=\partial_\rho f_1\,\mathbf e_\phi+\frac{f_2'(\rho)}{\rho}\mathbf k.
%\label{eq:fieldsparam}
%\end{equation}
\begin{equation}
\mathbf E=-\partial_\rho \phi(\rho)\,\mathbf e_\rho,
\qquad
\mathbf B=-\partial_\rho A_z(\rho)\,\mathbf e_\phi+\frac{f'(\rho)}{\rho}\mathbf e_z\,,
\label{eq:fieldsparam}
\end{equation}
with $2\pi f(\rho)=\Phi(\rho)$ magnetic flux.

Because the system is invariant under rotations around and translations along the $z$ axis, the operators $\hat L_z$ and $\hat p_z$ are conserved, and the wave function can be decomposed into sectors labeled by $(l,k_z)$.
%In units $\hbar=1$ and $2m=1$, 
The radial equation for $u=\psi_l(\rho)/\sqrt{\rho}$ takes the form
\begin{equation}
-\frac{\hbar^2}{2m}u''(\rho)+V_{l,k_z}(\rho)u(\rho)=E\,u(\rho),
\label{eq:radialeq}
\end{equation}
with
\begin{equation}
V_{l,k_z}(\rho)
=
\frac{\left(e f(\rho)-\hbar l\right)^2-\frac{\hbar^2}{4}}{2m\rho^2}
+e \phi(\rho)
+\frac{\left(\hbar k_z-eA_z(\rho)\right)^2}{2m}.
\label{eq:radialpotential}
\end{equation}

Fix now a partial wave $l$ and consider an arbitrary variation of $f(\rho)$. 
The radial Hamiltonian \eqref{eq:radialeq} is invariant, for all $k_z$, under the 
transformation
%\begin{eqnarray}
%f_2(\rho)&\to&\tilde f_2(\rho)=f_2(\rho)+\delta f_2(\rho),\\
% \nonumber f_0(\rho)&\to&\tilde f_0(\rho)
%=
%f_0(\rho)
%-
%\frac{2\delta f_2(\rho)\bigl(f_2(\rho)-l\bigr)+\delta f_2^2(\rho)}{\rho^2},
%\label{eq:partialsym}
%\end{eqnarray}
\begin{eqnarray}
f(\rho)&\to&\tilde f(\rho)=f(\rho)+\delta f(\rho),\\
 \nonumber \phi(\rho)&\to&\tilde \phi(\rho)
=
\phi(\rho)
-
\frac{2\delta f(\rho)\bigl(e f(\rho)-l\hbar\bigr)+e\delta f^2(\rho)}{2m\rho^2},
\label{eq:partialsym}
\end{eqnarray}
with arbitrary physically admissible $\delta f(\rho)$, while $A_z(\rho)$ remains unchanged. 
In contrast, the effective potentials in all other angular-momentum channels transform as
\begin{equation}
\tilde V_{l'}(\rho)=V_{l'}(\rho)+\frac{\hbar(l-l')e\,\delta f(\rho)}{m\rho^2}.
\label{eq:otherchannels}
\end{equation}
We therefore obtain an infinite-dimensional family of physically distinct electromagnetic backgrounds 
\begin{eqnarray}\label{def-field-transformation}
\tilde B_z &=& B_z+\dfrac{1}{\rho}\delta\! f' \\[2mm]
\tilde E_\rho &=& E_\rho+(ef-\hbar l)\frac{\rho\delta\! f'-2\delta\! f}{m\rho^3}\\
&+&\frac{e\delta\!f(\rho f'+\rho\delta\!f'-\delta\!f)}{m\rho^3} \nonumber
\end{eqnarray}
that are dynamically equivalent in one selected $L_z=l$ sector and inequivalent in all others. 

We refer to this property as a \emph{partial isophase symmetry} because 
not only the energy levels from the selected spectrum are invariant, but also
any wave packet 
with a definite $L_z=l$ undergoes identical evolution in all possible electromagnetic configurations \eqref{def-field-transformation}.

The remainder of this Letter develops this mechanism along the axially symmetric example above, then generalizes it. 
Since transformations \eqref{eq:partialsym} have a functional ambiguity, additional constraints
lead to rich possibilities presented in 
Secs.~\ref{sec:uniform}--\ref{sec:2channel}. Variation of the homogeneous part of the magnetic field naturally produces a rigid spectral flow of the Fock--Darwin oscillator (and brings some information beyond the harmonic background). Fixing the electric field, 
we get a one-parameter family of magnetic fields. Choosing \eqref{eq:otherchannels} to be a Darboux isospectral deformation, we 
extend our spectral control to two channels. Section~\ref{sec:inverse} shows that the symmetry is a genuinely new obstruction for the inverse spectral problem of the magnetic Schr\"odinger operator \cite{bellassoued2018inverse}, complementary to the classical Borg--Gel'fand--Levitan theory for a scalar potential; Sec.~\ref{sec:quasiclassical} is devoted to the quasiclassical counterpart of the partial isophase symmetry. Section~\ref{sec:general} then abstracts the construction to an arbitrary separable geometry, and Appendix~\ref{sec:slab} works it out in full for a Cartesian slab geometry that maps directly onto magnetic-barrier and velocity-filter devices already realized in two-dimensional electron gases \cite{matulis1994wavevector,nogaret2010electron}, which suggests possible applications.

\section{Uniform-field deformation and rigid spectral flow}
\label{sec:uniform}
\begin{figure}
    \centering
    \includegraphics[width=\linewidth]{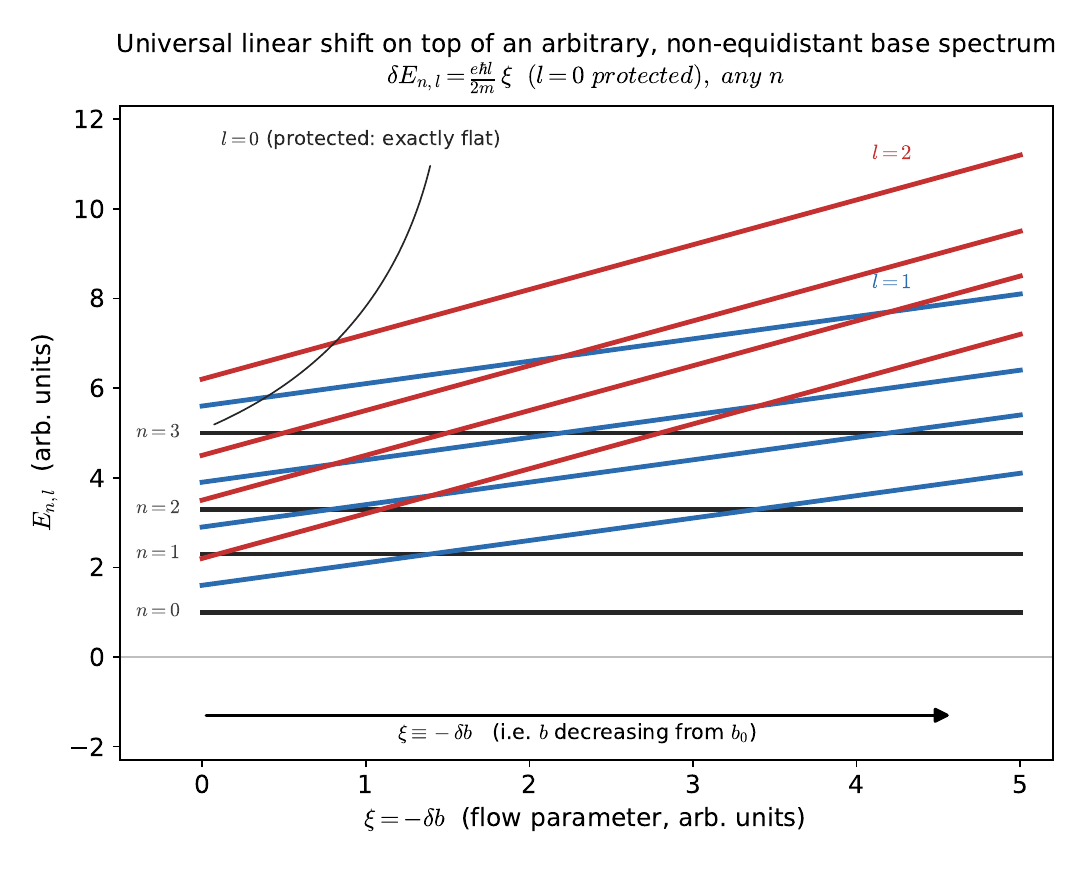}
    \caption{Spectral flow \eqref{fock-darwin-spectral-flow} for $l=0$ (protected), $1$, $2$ versus $\xi=-\delta b$, on top of a non-equidistant bare spectrum $E_{n,0}(b_0)$ (levels $n=0,\dots,3$). Every $n$ shifts linearly by $\delta E_{n,l}=e\hbar l\,\xi/2m$, whatever the bare spectrum; the $l=0$ levels stay flat.}
    \label{fig:rigid-levels}
\end{figure}

For an initially homogeneous magnetic field,
\begin{equation}
    f(\rho)=\frac{b\rho^2}{2},
    \qquad
    B_z^{(0)}=b ,
\end{equation}
choose
\begin{equation}
    \delta f(\rho)=\frac{\rho^2 \delta b}{2}.
\end{equation}
The transformation changes the magnetic field by
\begin{equation}
    \delta B_z=\delta b ,\label{fock-darvin-dB-field}
\end{equation}
and generates the compensating radial electric field
%\begin{equation}
%    \delta E_\rho
%    =
%    \frac{c\hbar}{e}\rho
%    \left(
%        \omega\delta\omega+\frac{\delta\omega^2}{2}
%    \right).
%\end{equation}
\begin{equation}
    \delta E_\rho
    =
    \frac{e\rho}{2m}
    \left(
        b\delta b+\frac{\delta b^2}{2}
    \right).
    \label{fock-darvin-dE-field}
\end{equation}
The protected \(L_z=l\) radial Hamiltonian is unchanged, 
whereas the other partial waves are shifted by constants,
\begin{equation}\label{fock-darwin-spectral-flow}
    \delta V_{l'}= \frac{e \hbar(l-l')}{2m}\delta b \,,\quad \delta E_{n,l'}=\delta V_{l'} ,
\end{equation}
%Consequently,
%\begin{equation}
%    \delta E_{n,l'}=(l-l')\delta\omega ,
%\end{equation}
and the radial eigenfunctions are unchanged.
Applying global gauge transformation $\phi(\rho)\to \phi(\rho)- \frac{\hbar l}{2m}\delta b$
we can always choose $l=0$ as a protected sector without losing generality. 
Note that the initial radial component of the electric field can be arbitrary. 
That is, the initial spectrum $E_{n,l'}$ and eigen-functions may not be analytically known. 
Nevertheless, the spectral flow (see Fig. \ref{fig:rigid-levels})
is given by \eqref{fock-darwin-spectral-flow} and wave functions in all channels are invariant.

\begin{figure}
    \centering
    \includegraphics[width=\linewidth]{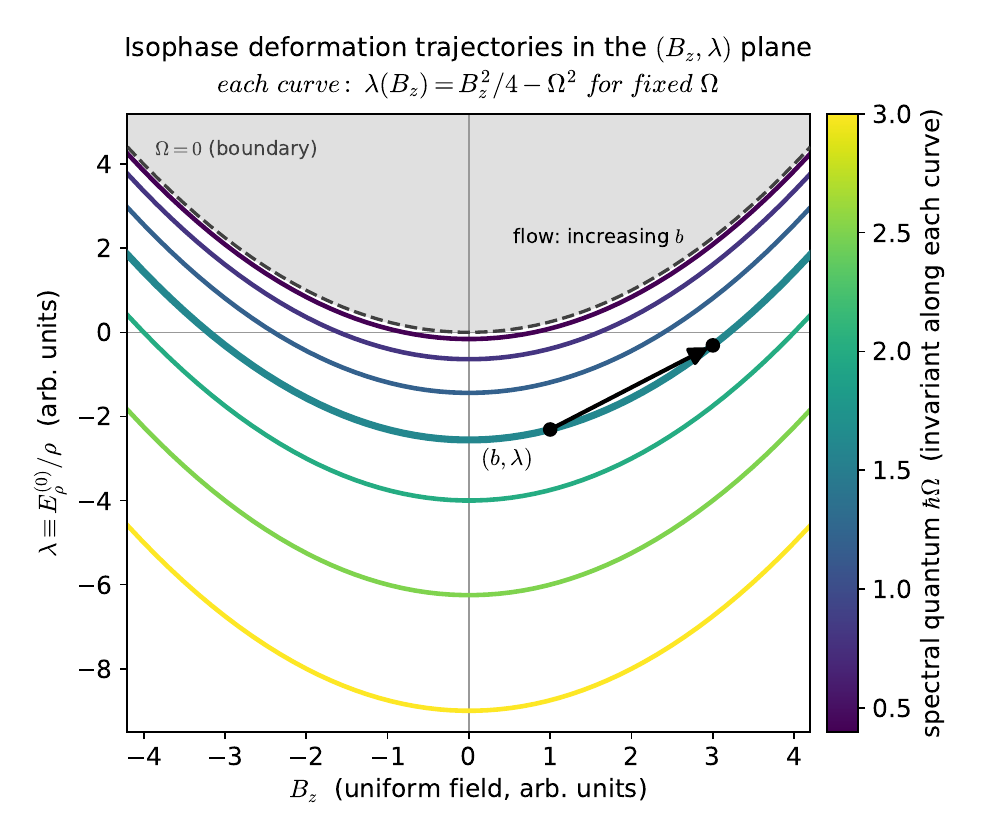}
    \caption{Trajectories of the transformation \eqref{fock-darvin-dB-field},\eqref{fock-darvin-dE-field} in the $(B_z,\lambda)$ plane, $\lambda\equiv E_\rho^{(0)}/\rho$. Along each parabola $\lambda=B_z^2/4-\Omega^2$ the spectral quantum $\hbar\Omega$ (color) is constant; the shaded region is $\Omega^2<0$. The arrow is one step $b\to b+\delta b$. For a non-linear background $E_\rho^{(0)}(\rho)$, $\lambda$ is the coefficient of its linear part, the only one changed by the flow.}
    \label{fig:FD-flow}
\end{figure}

When $E_\rho^{(0)}(\rho)=\lambda \rho$, family \eqref{fock-darvin-dB-field},\eqref{fock-darvin-dE-field}
generates a rigid spectral flow for the Fock-Darwin system \cite{fock1928bemerkung,darwin1931diamagnetism,drigho2017superintegrability}.
Transformation \eqref{fock-darvin-dB-field}, \eqref{fock-darvin-dE-field} gives a family of parabolas in the parameter space of the Fock-Darwin system (see Fig. \ref{fig:FD-flow}).
In this case, we have an exactly solvable $2D$ harmonic oscillator with the spectrum 
$E_{n,l}=(2n+|l|+1)\hbar \omega -\frac{1}{2}\hbar l\, \omega_c$.
The Larmor frequency $\omega_c=e B_z/m$ shifts the ground state energies of radial Hamiltonians
$E_{0,l}(B_z)$
accompanying 
variations of $B_z$. While the energy quantum ${\hbar \omega=\hbar \sqrt{\omega_c^2/4-e\lambda/m} }$ stays 
fixed under the flow \eqref{fock-darvin-dB-field},\eqref{fock-darvin-dE-field}
\begin{eqnarray}
   \sqrt{\frac{e^2(b+\delta b)^2}{4m^2}-\frac{e}{m}\left(\lambda+\frac{e\delta b}{4m}(2b+\delta b )\right)} \,\\
   =\sqrt{\frac{e^2b^2}{4m^2}-\frac{e\lambda}{m}} 
   \label{eq:FD-omega-invariance}
\end{eqnarray}

\section{Unmodified electric field}
\label{sec:unmodifiedE}
A particularly important case of symmetry appears in the case when we 
require $\delta E_\rho(\rho)=0$. 
In this case 
\[
\delta\phi=\frac{2\delta f(\rho)\bigl(ef(\rho)-\hbar l\bigr)+e\delta f^2(\rho)}{2m\rho^2}=C
\]
Thus we get \cite{pupasov2025propagators}
\[
e\delta f(\rho)=(\hbar l-ef(\rho))\pm\sqrt{(ef(\rho)-\hbar l)^2+2me\rho^2 C}\,,
\]
\[
\tilde B_z(\rho) = \frac{|ef(\rho)-\hbar l|B_z(\rho)+2mC}{\sqrt{(ef(\rho)-\hbar l)^2+2 m e\rho^2 C}}
\]
where the sign is chosen to provide a continuity of $\delta f(\rho)$ and $\tilde B_z(\rho)$.

Parameter $C$, $eC\geq0$, is responsible for a gauge symmetry (a constant shift of the scalar potential accompanied by a global phase transformation 
of the wave function phase $\exp(-iCt)$). If the initial electric field $\mathbf E=0$, we get a family of magnetic fields with 
isophase quantum dynamics in the $l$th OAM channel.

\section{Two-channel isospectrality}
\label{sec:2channel}
For $A_z(\rho)=0$ (and hence $B_\phi=0$), the partial-isophase construction can be extended to preserve one additional angular-momentum channel spectrally. Let $l$ denote the protected sector of the exact partial-isophase symmetry, and let $l_s\neq l$ be another fixed partial wave. 
%Since
%\begin{equation}
%\tilde V_{l_s}(\rho)-V_{l_s}(\rho)
%=
%\frac{2(l-l_s)\,\delta f_2(\rho)}{\rho^2},
%\label{eq:Vls-shift}
%\end{equation}
We may require deformation $\Delta \tilde V_{l_s}(\rho)$ 
to coincide with a standard isospectral Darboux transformation,
\begin{equation}
\tilde V_{l_s}(\rho)-V_{l_s}(\rho)
=
-2 \frac{\hbar^2}{2m}\frac{d^2}{d\rho^2}\ln W(\rho),
\label{eq:ls-darboux-condition}
\end{equation}
where $W(\rho)$ is chosen so that $V_{l_s}$ and $\tilde V_{l_s}$ are isospectral. This yields
\begin{equation}
\delta f(\rho)
=
\frac{\hbar\rho^2}{e(l_s-l)}\,
\frac{d^2}{d\rho^2}\ln W(\rho).
\label{eq:2channel-deltaf2}
\end{equation}
Hence, the electromagnetic deformation generated by \eqref{eq:2channel-deltaf2} preserves the $l$-channel exactly and makes the $l_s$ channel isospectral to its Darboux partner.

Explicit realizations are obtained by starting from an exactly solvable radial potential $V_{l_s}(\rho)$. In the general Crum-Darboux construction,
\begin{equation}
W(\rho)=\mathrm{Wr}\!\left[u_1(\rho,\alpha_1),\dots,u_N(\rho,\alpha_N)\right],
\label{eq:Wronskian}
\end{equation}
where the auxiliary functions satisfy
\begin{equation}
-\frac{\hbar^2}{2m}u_j''(\rho,\alpha_j)+V_{l_s}(\rho)u_j(\rho,\alpha_j)
=
\alpha_j\,u_j(\rho,\alpha_j),
\label{eq:auxiliary-radial}
\end{equation}
with factorization constants chosen below the ground-state energy, $\alpha_j<E_0^{(l_s)}$, so that the resulting Wronskian is nonsingular. The corresponding deformation \eqref{eq:2channel-deltaf2} then produces a family of electromagnetic backgrounds that is exactly invariant in the $l$ sector and spectrally equivalent in the $l_s$ sector.

Although Eq.~\eqref{eq:auxiliary-radial} is written for $k_z=0$, the conclusion extends to all longitudinal momenta. Indeed, for $A_z=0$ the $k_z$ dependence enters the radial Hamiltonian only through an additive constant, so $V_{l_s,k_z}$ and $\tilde V_{l_s,k_z}$ remain isospectral for arbitrary $k_z$.

An explicit realization with two Crum seeds below the ground state of the harmonic oscillator is worked out in Appendix~\ref{app:example}.

\section{Inverse spectral problem perspective}
\label{sec:inverse}
In standard literature on the inverse spectral problem for 3D Schr\"{o}dinger equation, electric and magnetic fields are 
considered separately. 
In the standard setting $H=-\Delta+U$ with separable variables, the single function $U$ enters all channels, so complete spectral data of one channel determine $U$, and with it every other channel. ``Complete'' means eigenvalues together with norming constants (or two spectra) \cite{borg1946umkehrung,gelfand1951determination,levitan1987inverse,poschel1987inverse}; for radial problems with a centrifugal term see \cite{chadan1989inverse,albeverio2007inverse}. Eigenvalues alone are not sufficient: the potentials with a prescribed spectrum form the isospectral (Crum--Darboux) manifold parametrized by the norming constants \cite{crum1955associated,cooper1995supersymmetry,poschel1987inverse}. 

Inverse results for the magnetic Schr\"odinger operator are usually formulated for boundary data \cite{bellassoued2018inverse,eskin1989inverse}, which in a separable geometry contain all angular-momentum channels at once.
If one considers ``spectroscopic'' information for magnetic backgrounds,
$\phi=\mathrm{const}$, Eq.~\eqref{eq:radialpotential} gives $(ef-\hbar l)^2$ in terms of $V_l$, so one channel fixes $f$ up to the choice of the branch of the square root and an additive constant (the parameter $C$ of Sec.~\ref{sec:unmodifiedE}). 

For the electromagnetic field the counting is different. A sector $(l,k_z)$ with complete data yields one function of $\rho$,
\begin{equation}
V_{l,k_z}=\frac{(ef-\hbar l)^2-\hbar^2/4}{2m\rho^2}+e\phi+\frac{(\hbar k_z-eA_z)^2}{2m},
\end{equation}
whereas the fields are described by three, $f$, $\phi$ and $A_z$. The partial isophase transformation \eqref{eq:partialsym} is precisely the description of the resulting ambiguity: for fixed $l$ (and all $k_z$) the data determine $A_z$ and the combination $V_l$, but the field is fixed only up to an arbitrary $\delta f(\rho)$. Additional sectors remove it algebraically. Two channels $l\neq l'$ give
\begin{eqnarray}
ef &=&\frac{\hbar(l+l')}{2}+\frac{m\rho^2\,(V_l-V_{l'})}{\hbar\,(l'-l)},\cr
eA_z &=& \frac{\hbar k_z}{2}-\frac{m\,(V_{l,k_z}-V_{l,0})}{\hbar k_z},
\label{eq:inverse-algebraic}
\end{eqnarray}
and $e\phi$ then follows from $V_l$. We therefore conclude that three sectors, e.g.\ $(l,0)$, $(l,k_z\neq0)$, $(l',0)$, are sufficient in principle, provided the potential of each channel can be reconstructed from its data; the exponent $\nu=|l-f(0)e/\hbar|$ of the centrifugal term is encoded in the asymptotics of the levels and norming constants. Equation~\eqref{eq:inverse-algebraic} also explains Eq.~\eqref{eq:otherchannels}: the deformation of channel $l'$ is proportional to $l-l'$. A single channel is not sufficient even with complete data, and the corresponding non-uniqueness is infinite-dimensional.
We do not address the stability of the reconstruction, the sufficiency of finitely many levels, or whether isospectrality can be imposed simultaneously in more than one unprotected channel.

\section{Quasi-classical limit}
\label{sec:quasiclassical}

In the classical limit, $\hbar\to0$, $l\to\infty$ with $L=\hbar l$ fixed, the canonical angular momentum $L$ is conserved and the radial motion is governed by
\begin{equation}
H_L=\frac{p_\rho^2}{2m}+V_L(\rho),\qquad V_L(\rho)=\frac{(L-ef)^2}{2m\rho^2}+e\phi ,
\label{eq:classicalV}
\end{equation}
the classical counterpart of Eq.~\eqref{eq:radialpotential} (the $z$ motion is unaffected, since $A_z$ is not changed). The fields transform as in the quantum case, Eq.~\eqref{eq:partialsym} with $\hbar l\to L_0$. Then $V_{L_0}$ is invariant, so the whole radial dynamics $(\rho(t),p_\rho(t))$ at $L=L_0$ is unchanged, and for $L_0$ the orbits fill the same annulus $\rho_{\min}\le\rho\le\rho_{\max}$ determined by $E=V_{L_0}(\rho)$. Comparing the Hamiltonians at the same $(\rho,p_\rho,L)$, the $\delta f^2$ terms cancel and
\begin{equation}
\tilde H-H=-\,\Omega_\delta(\rho)\,(L-L_0),\qquad \Omega_\delta(\rho)=\frac{e\,\delta f(\rho)}{m\rho^2}.
\label{eq:differential-rotation}
\end{equation}
The same identity holds for the quantum Hamiltonian, $\tilde H-H=-\Omega_\delta(\rho)(\hat L_z-\hbar l)$, because $\Omega_\delta(\rho)$ commutes with $\hat L_z$; it reproduces Eq.~\eqref{eq:otherchannels}. The transformation thus acts as a \emph{differential rotation} with angular velocity $\Omega_\delta(\rho)$, which vanishes identically in the protected sector.
\begin{figure*}
    \centering
    \includegraphics[width=\textwidth]{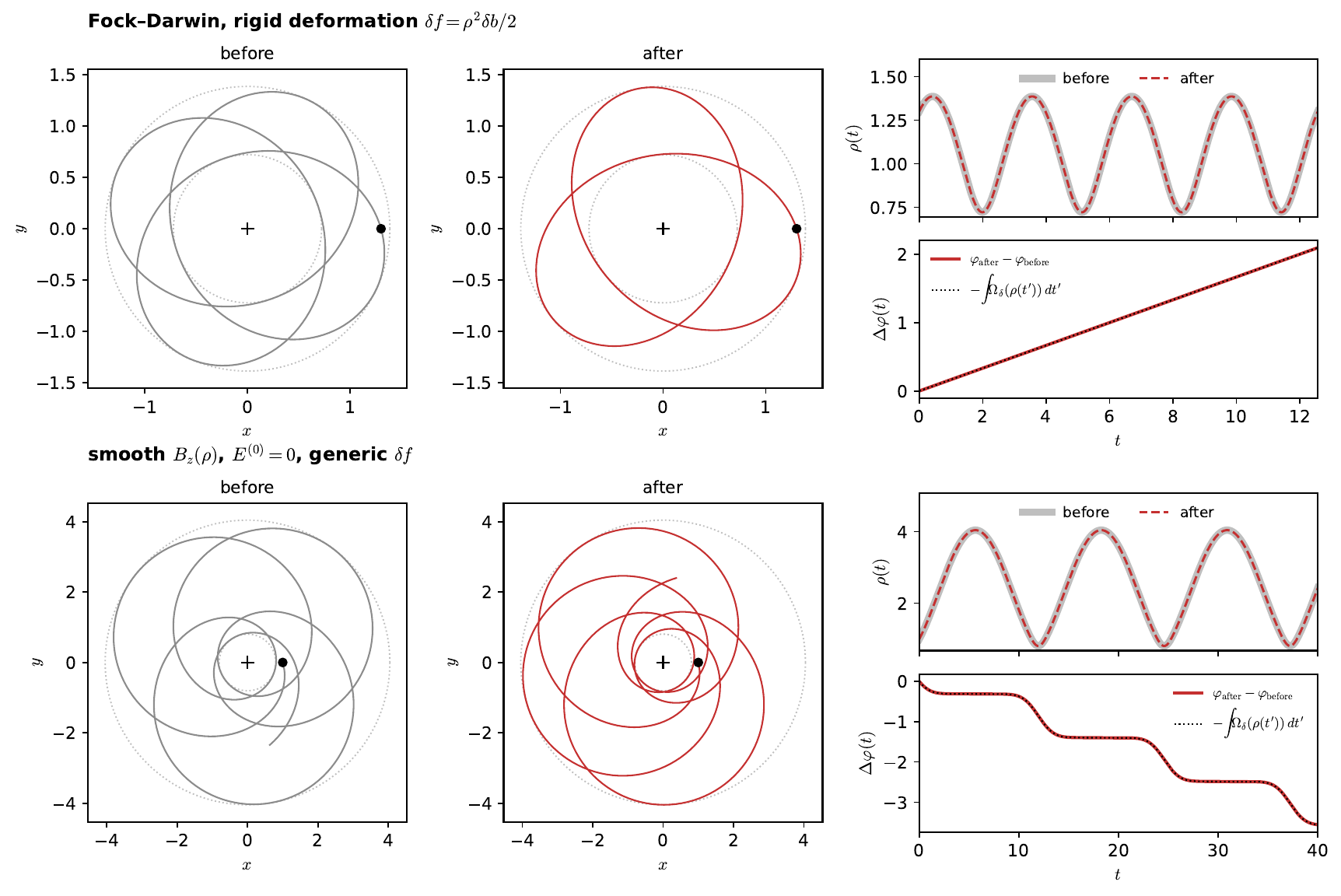}
    \caption{Classical orbits with $L=L_0$ before and after the transformation ($e=m=1$), obtained by integrating Newton's equations with the Lorentz force; the dot marks the initial point, dotted circles are the turning radii. Top: Fock--Darwin system, $b=1$, $\lambda=-0.75$ ($\omega_+/\omega_-=3$, closed four-lobe orbit), $\delta f=\rho^2\delta b/2$, $\delta b=-1/3$ ($\omega_+/\omega_-=2$, closed three-lobe orbit). Bottom: $B_z=0.6+1.4\,e^{-\rho^2/1.3^2}$, $E^{(0)}=0$, $\delta f=0.5\,\rho^2e^{-\rho^2/2}$. Right: $\rho(t)$ is identical before and after (to $10^{-9}$), while $\varphi_{\rm after}-\varphi_{\rm before}=-\int\Omega_\delta(\rho(t'))\,dt'$ (dotted) grows linearly (rigid rotation) in the quadratic case and by steps (differential rotation) otherwise.}
    \label{fig:classical-orbits2}
\end{figure*}
The real-space orbit, however, is not invariant. Since $m\rho^2\dot\varphi=L-ef$, Eq.~\eqref{eq:differential-rotation} gives $\dot{\tilde\varphi}=\dot\varphi-\Omega_\delta(\rho)$, and because $\rho(t)$ is the same,
\begin{equation}
\tilde\varphi(t)=\varphi(t)-\int_0^t\Omega_\delta\big(\rho(t')\big)\,dt' .
\label{eq:azimuth-shift}
\end{equation}
The velocity, the kinetic energy and the Lorentz force along the orbit change; only the combination entering the radial equation, $m(\ddot\rho-\rho\dot\varphi^2)=e(E_\rho+\rho\dot\varphi B_z)$, is invariant, although each of its terms is modified. The partial isophase property is therefore a property of the reduced radial dynamics rather than of the trajectory in the plane. It is not restricted to circular orbits: a circular orbit keeps its radius ($V_L'=0$) but is traversed with a different angular velocity.

For the Fock--Darwin system, $f=b\rho^2/2$, $E_\rho^{(0)}=\lambda\rho$, the orbit is a superposition of two counter-rotating circular modes with frequencies $\omega_\pm=\omega\pm\omega_c/2$, and $\rho(t)$ oscillates with frequency $2\omega$. The orbit is closed if and only if $\omega_+/\omega_-$ is rational. For the uniform-field deformation, $\Omega_\delta=e\,\delta b/2m$ is constant: Eq.~\eqref{eq:azimuth-shift} is a rigid rotation, $\omega$ (and $\rho(t)$) is fixed, which is the classical counterpart of the invariance of $\hbar\omega$ in Eqs.~\eqref{fock-darvin-dB-field},\eqref{fock-darvin-dE-field}, while $\omega_\pm\to\omega_\pm\pm\Omega_\delta$. The shape of a single loop is preserved, but the closure condition changes (Fig.~\ref{fig:classical-orbits2}, top). For a generic $\delta f$ (Fig.~\ref{fig:classical-orbits2}, bottom) the rotation is differential: the azimuth is shifted in steps, mainly where $\delta f$ is non-negligible, the orbit in the plane is distorted, and the radial oscillations are not isochronous; nevertheless $\rho(t)$ coincides exactly.

Thus the classical dynamics within a family of partially isophase backgrounds is consistent with the quantum picture. 
The density of a wave packet with a sharp $l$, built from radial states, forms a ring of pulsating radius, and its evolution is the same before and after the transformation. All observables that are functions of $(\rho,p_\rho,\hat L_z)$ coincide. Note that gauge-invariant kinetic quantities transform as follows: the current $j_\varphi=(\hbar l-ef)|\psi|^2/m\rho$, the kinetic energy and the orbital magnetic moment change, being the quantum counterparts of the shift of $\dot\varphi$. In the protected channel the configurations are therefore indistinguishable by the spectrum and the density, but distinguishable by currents and magnetization. For a packet with a spread $\Delta l$ the deviation from identical evolution is controlled by Eq.~\eqref{eq:differential-rotation} and grows with $\Delta l$. Finally, the Bohr--Sommerfeld condition $\oint p_\rho\,d\rho=2\pi\hbar(n+\gamma)$ depends only on $V_l$ and gives the same levels in the protected channel. 

\section{General scheme: separable coordinates and sector protection}
\label{sec:general}

The construction of Secs.~\ref{sec:Intro}--\ref{sec:2channel} rests on a structural feature of the cylindrical problem that is worth isolating, since it can be extended to other separable geometries.

Suppose the electromagnetic background is invariant under a set of commuting continuous symmetries (translations and/or rotations) labeled by $a=1,\dots,k$, so that after separation of variables in the remaining coordinate $\xi$ the wave function carries good quantum numbers $n_a$ (linear and/or angular momenta), and the vector potential admits, in a suitable gauge, one component along each symmetry direction. It is convenient to write the minimal-coupling combination along direction $a$ as $\hbar n_a-e f_a(\xi)$, where $f_a(\xi)=A_a(\xi)$ for a translational direction and $f_a(\xi)=\rho A_\varphi(\xi)$ (the flux function) for rotation about the $\xi$-axis. The effective one-dimensional potential in the sector $\{n_a\}$ then takes the general form
\begin{equation}
V_{\{n_a\}}(\xi)=\sum_a \frac{c_a(\xi)}{2m}\bigl(\hbar n_a-e f_a(\xi)\bigr)^2+e\phi(\xi)+Q(\xi),
\label{eq:generalV}
\end{equation}
where $c_a(\xi)$ is a direction-dependent weight ($c_a=1/\rho^2$ for rotation about $\xi=\rho$, $c_a=1$ for translation), and $Q(\xi)$ is a measure-induced quantum correction that is present only for curved (rotational) directions -- cf.\ the $-\hbar^2/(8m\rho^2)$ term hidden in Eq.~\eqref{eq:radialpotential} -- and vanishes identically for flat (translational) ones. Equation~\eqref{eq:radialpotential} is recovered for $k=2$, with $a=1$ the rotational direction ($c_1=1/\rho^2$, $f_1=f$) and $a=2$ the translational direction along $z$ ($c_2=1$, $f_2=A_z$, absorbed with a minus sign into the definition used in Sec.~\ref{sec:Intro}).

Fix one direction $a_0$ and one value $n_{a_0}=n_{a_0}^0$ to be protected. Deform only the corresponding flux function, $f_{a_0}\to\tilde f_{a_0}=f_{a_0}+\delta f_{a_0}$, leaving every $f_a$ with $a\neq a_0$ untouched. Demanding that $V_{\{n_a\}}(\xi)$ with $n_{a_0}=n_{a_0}^0$ be left invariant as a function of $\xi$, for arbitrary values of the remaining quantum numbers, fixes the compensating scalar potential
\begin{equation}
\delta\phi(\xi)=-c_{a_0}(\xi)\,\frac{2\delta f_{a_0}(\xi)\bigl(ef_{a_0}(\xi)-\hbar n_{a_0}^0\bigr)+e\,\delta f_{a_0}^2(\xi)}{2m},
\label{eq:generalcompensation}
\end{equation}
which reduces to Eq.~\eqref{eq:partialsym} for $c_{a_0}=1/\rho^2$, $n_{a_0}^0=l$. Every other value $n_{a_0}'\neq n_{a_0}^0$ along the same direction (with all other $n_a$ unchanged) then shifts by
\begin{equation}
\Delta V_{n_{a_0}'}(\xi)=c_{a_0}(\xi)\,\frac{e\hbar\bigl(n_{a_0}^0-n_{a_0}'\bigr)}{m}\,\delta f_{a_0}(\xi),
\label{eq:generalshift}
\end{equation}
which reduces to Eq.~\eqref{eq:otherchannels}. Both reductions follow by direct substitution of Eq.~\eqref{eq:generalcompensation} into Eq.~\eqref{eq:generalV} and are exact, not merely first-order in $\delta f_{a_0}$.

Two structural remarks follow immediately from Eqs.~\eqref{eq:generalV}--\eqref{eq:generalshift}. First, the construction of Sec.~\ref{sec:Intro} can equally well be applied to $A_z(\rho)$ rather than to $f(\rho)$, protecting a chosen $k_z^0$ for all $l$ instead of a chosen $l$ for all $k_z$; since $\delta\phi(\xi)$ is a single function, only one such direction can be protected in full, and protecting a single point $(l_0,k_z^0)$ in the space of quantum numbers by splitting the freedom between $\delta f$ and $\delta A_z$ is possible only at the cost of the ``for all'' quantifier. Second, whenever $k\ge 2$ independent symmetry directions are available, the same trade-off recurs: the single scalar potential can protect one direction completely, or several directions only at isolated points of their joint spectrum. Appendix~\ref{sec:slab} illustrates the $k=2$ case explicitly, with both directions translational.
The Cartesian slab geometry is presented in Appendix~\ref{sec:slab}.

\section{Conclusion}
\label{sec:conclusion}

We have identified a family of electromagnetic deformations, a partial isophase symmetry, that leave the quantum dynamics in one selected separable sector exactly unchanged while modifying the physical field itself, complementary to ordinary gauge invariance. Its general mechanism (Sec.~\ref{sec:general}), its axially symmetric realizations (Secs.~\ref{sec:Intro}--\ref{sec:2channel}) and its Cartesian-slab realization (Appendix~\ref{sec:slab}) were worked out above.
The two-channel isospectrality presented in Sec.~\ref{sec:2channel} seems very interesting for further mathematical studies, since usually Darboux transformations of an effective radial Hamiltonian destroy analytical structure in other sectors (except for highly symmetric shape-invariant potentials). 
Ambiguities in spectrum-to-field inversion clarify a set of sufficient spectral data for the inverse 3D spectral problem in a mixed electromagnetic background (Sec.~\ref{sec:inverse}). In the classical limit (Sec.~\ref{sec:quasiclassical}) we found an interesting interplay between the angular and radial dynamics of 
individual trajectories. 
Our preliminary results also show that a similar symmetry manifests itself for non-stationary electromagnetic fields directly in the time-dependent 
Schr\"odinger equation, and in the Schr\"odinger--Poisson equation, where the wave packet describes a beam of identical particles in the mean-field approximation.

\begin{acknowledgments}
The author is grateful to Professor P. O. Kazinsky for valuable comments and suggestions. This work was presented at the seminar of the Department of Mathematics (PPGMPA), Universidade Federal de Santa Catarina. The author also thanks Ivan Pontual Costa e Silva and Vladimir Pestov for organizing this seminar.
\end{acknowledgments}

\appendix

\section{Explicit example of the two-channel isospectrality}
\label{app:example}

We take $l_s=0$, $l=1$ and a seed background consisting of a uniform field $B_z=\omega$ (in units of $\hbar/e$) and an Aharonov--Bohm flux $\mu=1/2$, i.e.\ $F(\rho)\equiv ef/\hbar=\mu+\omega\rho^2/2$, $\phi=0$; we use units $\hbar^2/2m=1$. Since $|l-\mu|=1/2$ in both channels, the centrifugal term in Eq.~\eqref{eq:radialpotential} vanishes, $V_0=\omega/2+\omega^2\rho^2/4$, $V_1=-\omega/2+\omega^2\rho^2/4$, and both channels reduce to the oscillator on the half-line with $u(0)=0$, $E_n^{(0)}=2\omega(n+1)$, $E_n^{(1)}=\omega(2n+1)$; the AB flux is what makes the origin regular for $l=0$ and $l=1$ simultaneously and allows oscillator seed functions to be used. The two Crum seeds of the $l_s=0$ channel, at energies $E_2<E_1<E_0^{(l_s)}=2\omega$, are $u_1=\rho\,e^{-\omega\rho^2/4}\,{}_1F_1\!\left(1-\tfrac{E_1}{2\omega};\tfrac32;\tfrac{\omega\rho^2}{2}\right)$, regular at the origin, and $u_2=e^{-\omega\rho^2/4}\,U\!\left(\tfrac12-\tfrac{E_2}{2\omega};\tfrac12;\tfrac{\omega\rho^2}{2}\right)$, decaying at infinity; we use $\omega=1$, $E_1=\omega/2$, $E_2=-3\omega/2$. Both seeds are nodeless, $W(0)<0$, $W'=(E_1-E_2)u_1u_2>0$ and $W\to0^-$, so $W=\mathrm{Wr}[u_1,u_2]<0$ for all $\rho>0$. Equation~\eqref{eq:2channel-deltaf2} then gives $\delta F=-\rho^2(\ln W)''$ (with $\delta F=e\,\delta f/\hbar$), and the scalar potential follows from Eq.~\eqref{eq:partialsym}, $e\tilde\phi=-[2\delta F(F-l)+\delta F^2]/\rho^2$.

\begin{figure*}
    \centering
    \includegraphics[width=\textwidth]{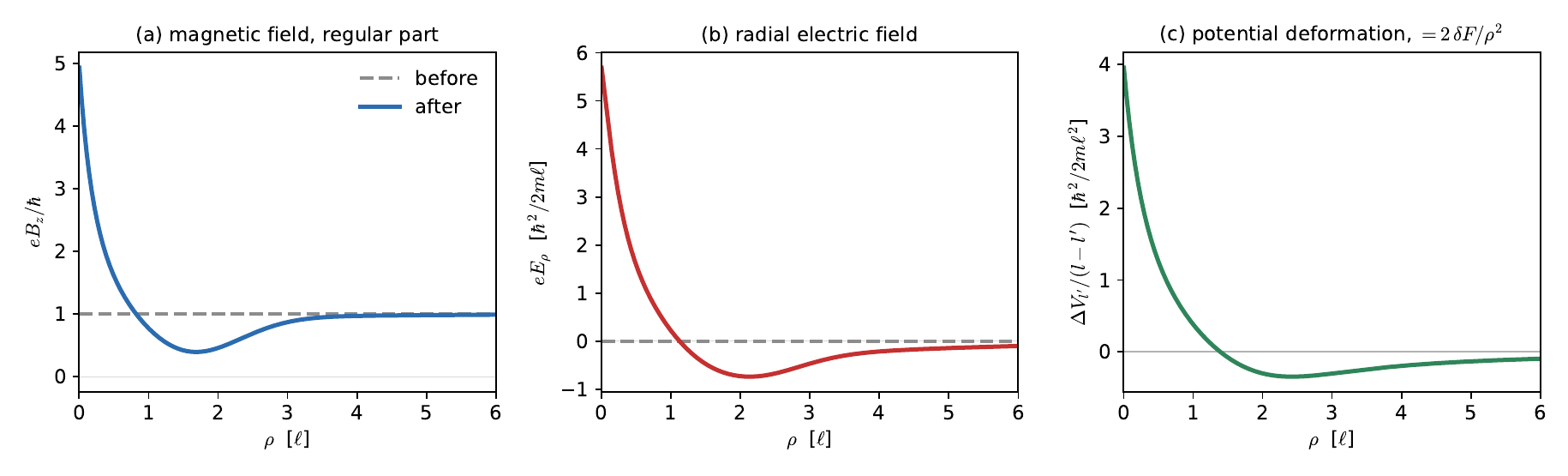}
    \caption{Regular parts of the fields before and after the two-channel deformation ($l_s=0$, $l=1$, $\mu=1/2$, $\omega=1$, $E_1=\omega/2$, $E_2=-3\omega/2$; $\hbar^2/2m=1$): (a) $B_z$, (b) $E_\rho$, (c) universal potential deformation $\Delta V_{l'}/(l-l')$. The AB flux $\mu=1/2$ is unchanged.}
    \label{fig:ex-fields}
\end{figure*}

\begin{figure*}
    \centering
    \includegraphics[width=\textwidth]{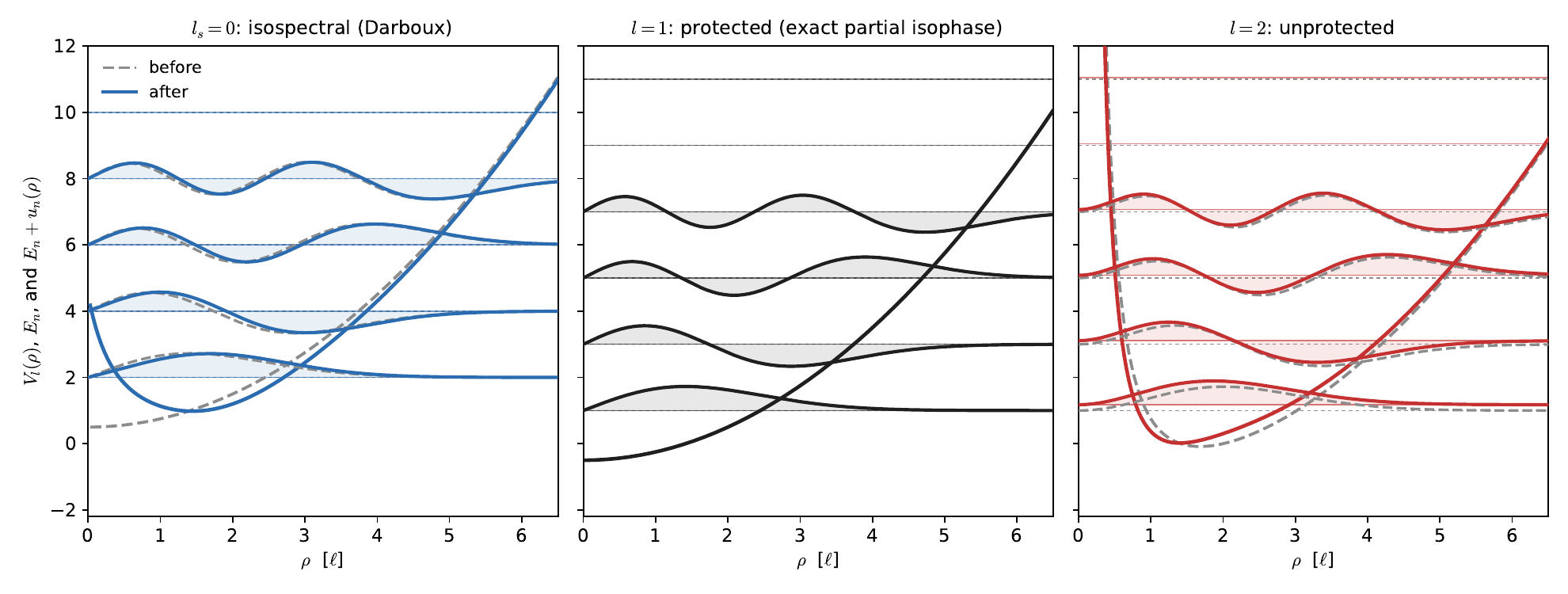}
    \caption{Potentials, levels and eigenfunctions $E_n+u_n(\rho)$ before (dashed) and after (solid) the deformation for the isospectral channel $l_s=0$, the protected channel $l=1$ (unchanged) and the unprotected channel $l=2$.}
    \label{fig:ex-channels}
\end{figure*}

\begin{figure*}
    \centering
    \includegraphics[width=0.85\textwidth]{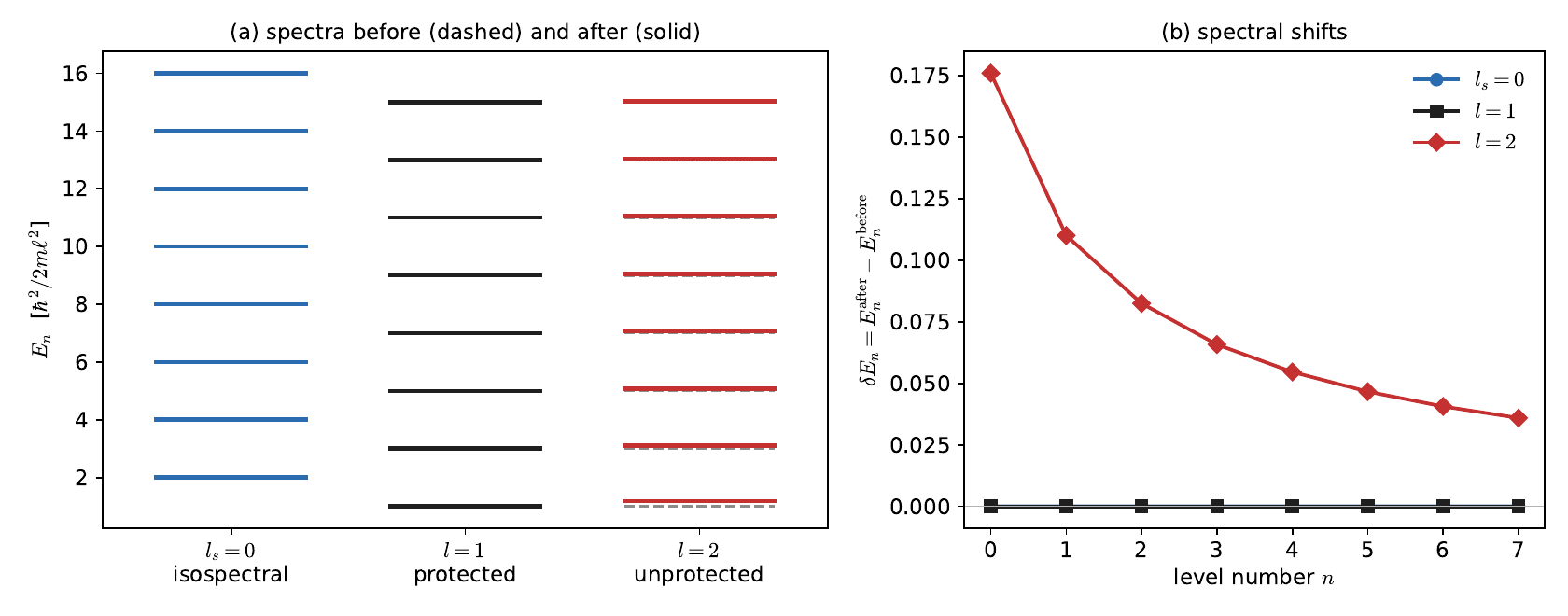}
    \caption{(a) Lowest levels of the three channels before (dashed) and after (solid); (b) shifts $\delta E_n$: zero for $l_s=0$ and $l=1$, positive and decreasing with $n$ for $l=2$.}
    \label{fig:ex-spectrum}
\end{figure*}

The AB flux is unchanged, $\delta F(0)=0$, so only the regular parts of the fields are modified (Fig.~\ref{fig:ex-fields}). The deformation is finite and localized: $\delta F\to-2$ at large $\rho$ (a net flux of $-2$ flux quanta), $\tilde B_z\to\omega$, and $e\tilde\phi\to2\omega$, a constant that keeps the $l_s=0$ channel exactly isospectral; at the origin the fields are finite but contain a term linear in $\rho$ ($\tilde B_z(0)\approx5$, $e\tilde E_\rho(0^+)\approx5.7$). By Eq.~\eqref{eq:otherchannels} all potential deformations are given by one universal function, $\Delta V_{l'}/(l-l')=2\,\delta F/\rho^2=-2(\ln W)''$ [Fig.~\ref{fig:ex-fields}(c)]. Numerical diagonalization of $-u''+Vu=Eu$ (Fig.~\ref{fig:ex-channels}) confirms that the $l=1$ potential, spectrum and eigenfunctions are unchanged; the $l_s=0$ channel keeps all levels $2\omega(n+1)$ while its eigenfunctions $\tilde u_n=\mathrm{Wr}[u_1,u_2,u_n]/\mathrm{Wr}[u_1,u_2]$ are visibly deformed; and the unprotected channel $l'=2$, degenerate with $l=1$ before the deformation, is shifted, $\delta E_n\simeq0.18,\,0.11,\,0.08,\,0.07,\,0.05,\dots$ (Fig.~\ref{fig:ex-spectrum}).

\section{Cartesian slab geometry}
\label{sec:slab}

\subsection{Setup}

As a first application of Sec.~\ref{sec:general} beyond cylindrical symmetry, consider a background invariant under translations along $y$ and $z$, with all nontrivial dependence carried by a single Cartesian coordinate $x$ (a ``slab''). In the gauge $A_x=0$,
\begin{equation}
\phi=\phi(x),\qquad \mathbf A=A_y(x)\,\mathbf e_y+A_z(x)\,\mathbf e_z,
\label{eq:slabpot}
\end{equation}
so that
\begin{equation}
\mathbf B=-A_z'(x)\,\mathbf e_y+A_y'(x)\,\mathbf e_z,\qquad \mathbf E=-\phi'(x)\,\mathbf e_x.
\label{eq:slabfields}
\end{equation}
Here both symmetry directions are translational, $c_y=c_z=1$ and $Q=0$ identically: unlike Eq.~\eqref{eq:radialpotential}, there is no measure-induced correction and no coordinate singularity at $x=0$.

\subsection{Reduction and sector protection}

With $p_y,p_z$ conserved and $\Psi=e^{i(k_yy+k_zz)}\chi(x)e^{-iEt/\hbar}$, Eq.~\eqref{eq:Sch3D} reduces to
\begin{equation}
-\frac{\hbar^2}{2m}\chi''(x)+V_{k_y,k_z}(x)\chi(x)=E\,\chi(x),
\label{eq:slabradial}
\end{equation}
\begin{equation}
V_{k_y,k_z}(x)=\frac{(\hbar k_y-eA_y(x))^2}{2m}+\frac{(\hbar k_z-eA_z(x))^2}{2m}+e\phi(x),
\label{eq:slabpotential}
\end{equation}
the $k=2$, translational instance of Eq.~\eqref{eq:generalV}. Applying Eqs.~\eqref{eq:generalcompensation}--\eqref{eq:generalshift} with $a_0=y$, $c_{a_0}=1$, $n_{a_0}^0=k_y^0$ gives the protecting deformation
\begin{equation}
\delta\phi(x)=\frac{2\delta A_y(x)\bigl(\hbar k_y^0-eA_y(x)\bigr)-e\,\delta A_y^2(x)}{2m},
\label{eq:slabtrick}
\end{equation}
\begin{equation}
\Delta V_{k_y'}(x)=\frac{e\hbar(k_y^0-k_y')}{m}\,\delta A_y(x),
\label{eq:slabshift}
\end{equation}
while $A_z(x)$ is left untouched (it can be protected in exactly the same way instead, protecting a chosen $k_z^0$ for all $k_y$; see the remark at the end of Sec.~\ref{sec:general}). The magnetic field changes by
\begin{equation}
\tilde B_z(x)=B_z(x)+\delta A_y'(x).
\label{eq:slabB}
\end{equation}

\subsection{Rigid spectral flow of a magnetically confined channel}

Take the Landau gauge $A_y(x)=bx$ ($B_z^{(0)}=b$) together with a background field $E_x^{(0)}(x)=\lambda x$, i.e.\ $\phi(x)=-\lambda x^2/2$; this is the standard model of a quantum wire in a perpendicular magnetic field with parabolic electrostatic confinement. Protecting $k_y^0=0$ (any other value can always be brought to $0$ by the same global gauge shift used in Sec.~\ref{sec:uniform}), Eq.~\eqref{eq:slabtrick} with $\delta A_y(x)=x\,\delta b$ gives $\delta\phi(x)=-\tfrac{e x^2}{2m}\bigl[(b+\delta b)^2-b^2\bigr]$. Adding this to the deformed kinetic term of Eq.~\eqref{eq:slabpotential},
\begin{eqnarray}
\tilde V_{0,k_z}(x)&=&\frac{e^2(b+\delta b)^2x^2}{2m}-\frac{e\lambda x^2}{2}-\frac{e^2x^2}{2m}\Bigl[(b+\delta b)^2-b^2\Bigr]\nonumber\\[1mm]
&&+\,\frac{(\hbar k_z-eA_z)^2}{2m}
=V_{0,k_z}(x),
\end{eqnarray}
the quadratic-in-$x$ deformation cancels identically, confirming Eq.~\eqref{eq:generalcompensation} by direct computation. The protected channel is a harmonic oscillator of frequency $\Omega^2=\omega_c^2-e\lambda/m$, $\omega_c=eb/m$, exactly as in Eq.~\eqref{eq:FD-omega-invariance}. For $k_y'\neq0$, Eq.~\eqref{eq:slabshift} gives $\Delta V_{k_y'}(x)=-\tfrac{e\hbar k_y'}{m}\delta b\,x$: a term linear in $x$, which only translates the center of the parabola without changing its curvature. Hence $\Omega$ is common to every channel, before and after the deformation -- a rigid spectral flow of the same kind as in Sec.~\ref{sec:uniform}, but for a single, non-degenerate transverse mode rather than a full Landau level.

\subsection{Unmodified electric field}

Requiring $\delta E_x(x)=0$ in Eq.~\eqref{eq:slabtrick} forces $\delta\phi=C=\mathrm{const}$. Writing $h(x)\equiv\hbar k_y^0-eA_y(x)$ (so $h'=-eB_z$), the continuous-at-$C=0$ solution is
\begin{equation}
e\,\delta A_y(x)=h(x)-\sqrt{h(x)^2-2meC},
\label{eq:slabC}
\end{equation}
and the deformed field follows in closed form,
\begin{equation}
\tilde B_z(x)=\frac{|h(x)|\,B_z(x)}{\sqrt{h(x)^2-2meC}},
\label{eq:slabCB}
\end{equation}
structurally identical to the cylindrical result of Sec.~\ref{sec:unmodifiedE}, but without the metric factors of $\rho$.

\subsection{Physical realization: magnetic barriers in a two-dimensional electron gas}

The geometry of Eqs.~\eqref{eq:slabpot}--\eqref{eq:slabpotential} is typical for \emph{magnetic barriers} in a two-dimensional electron gas (2DEG), realized by depositing a ferromagnetic or current-carrying stripe on a semiconductor heterostructure so as to produce a perpendicular field $B_z(x)$ that is localized across the transport direction $x$ and translation invariant along $y$ \cite{matulis1994wavevector,nogaret2010electron}. There, $k_y$ is the conserved transverse momentum (equivalently, the angle of incidence of an electron on the barrier), and Eq.~\eqref{eq:slabpotential} is exactly the effective 1D scattering potential seen by each incident channel. It is well established in this literature that an inhomogeneous $B_z(x)$ acts as a wavevector-dependent, rather than electrostatic, barrier: transmission depends sensitively on $k_y$, which is the basis of proposed velocity- and angle-filtering devices and of magnetic focusing schemes \cite{matulis1994wavevector,nogaret2010electron}.

Equations~\eqref{eq:slabtrick}--\eqref{eq:slabB} show that two magnetic barriers with visibly different profiles $B_z(x)\neq\tilde B_z(x)$---produced, say, by differently shaped ferromagnetic stripes - can be tuned, through an accompanying electrostatic gate voltage that realizes $\phi\to\phi+\delta\phi$, to be exactly indistinguishable in transmission and phase for one selected incidence channel $k_y^0$, while differing for every other channel. Since split gates alongside the ferromagnetic stripe are already a standard part of these devices \cite{nogaret2010electron}, no new fabrication step is required, only a specific choice of gate voltage as a function of $B_z(x)$ and $\tilde B_z(x)$ via Eq.~\eqref{eq:slabtrick}. This suggests two concrete uses: (i) an experimental test of the present construction, by comparing transmission through two differently fabricated barriers, tuned this way, at normal and oblique incidence; and (ii) a barrier that can be reshaped (e.g.\ to optimize some other figure of merit) while leaving the response of a fixed injection channel unaffected, so that off-channel changes serve as a built-in diagnostic of the deformation itself.

\subsection{Numerical study: a resonant magnetic barrier}
\label{sec:numerics}

To illustrate the construction quantitatively we consider transmission through a purely magnetic barrier in an infinite (in $y$) 2DEG strip, where $k_y$ is a good quantum number, and a deformation that protects one incidence channel $k_{y1}\equiv k_y^0$. Units are $\hbar=m=e=1$, with a length scale $\ell$ (see below).

\subsubsection{Barrier profile and deformation family}

The field of one edge of a magnetized stripe of thickness $H$ at depth $d$ below the 2DEG is $B_1(x)=b\ln\frac{x^2+d^2}{x^2+(d+H)^2}$, with the closed-form vector potential
\begin{eqnarray}
A_1(x)&=&b\Bigl\{x\ln\tfrac{x^2+d^2}{x^2+(d+H)^2}+2d\arctan\tfrac{x}{d}\nonumber\\
&&-2(d+H)\arctan\tfrac{x}{d+H}\Bigr\}.
\end{eqnarray}
A stripe of length $L_x$ is $B_1(x+L_x/2)-B_1(x-L_x/2)$ (zero net flux, so that $A_y\to0$ at $|x|\to\infty$). A single stripe gives a smooth non-resonant barrier. To obtain resonances we use two identical stripes with centers at $x=\pm g/2$; the region between them is a well in $V_{k_y}(x)$, and the structure acts as a Fabry--Perot resonator. We use $b=0.18$, $d=1$, $H=1.5$, $L_x=3$, $g=6$. For a GaAs 2DEG ($m^*=0.067\,m_e$) and $\ell=50$~nm this corresponds to an energy unit $\hbar^2/m^*\ell^2\approx0.455$~meV and a field unit $\hbar/e\ell^2\approx0.263$~T. For this profile $A_y(x)$ is even and $B_z(x)$ is odd.

\begin{figure*}
    \centering
    \includegraphics[width=\textwidth]{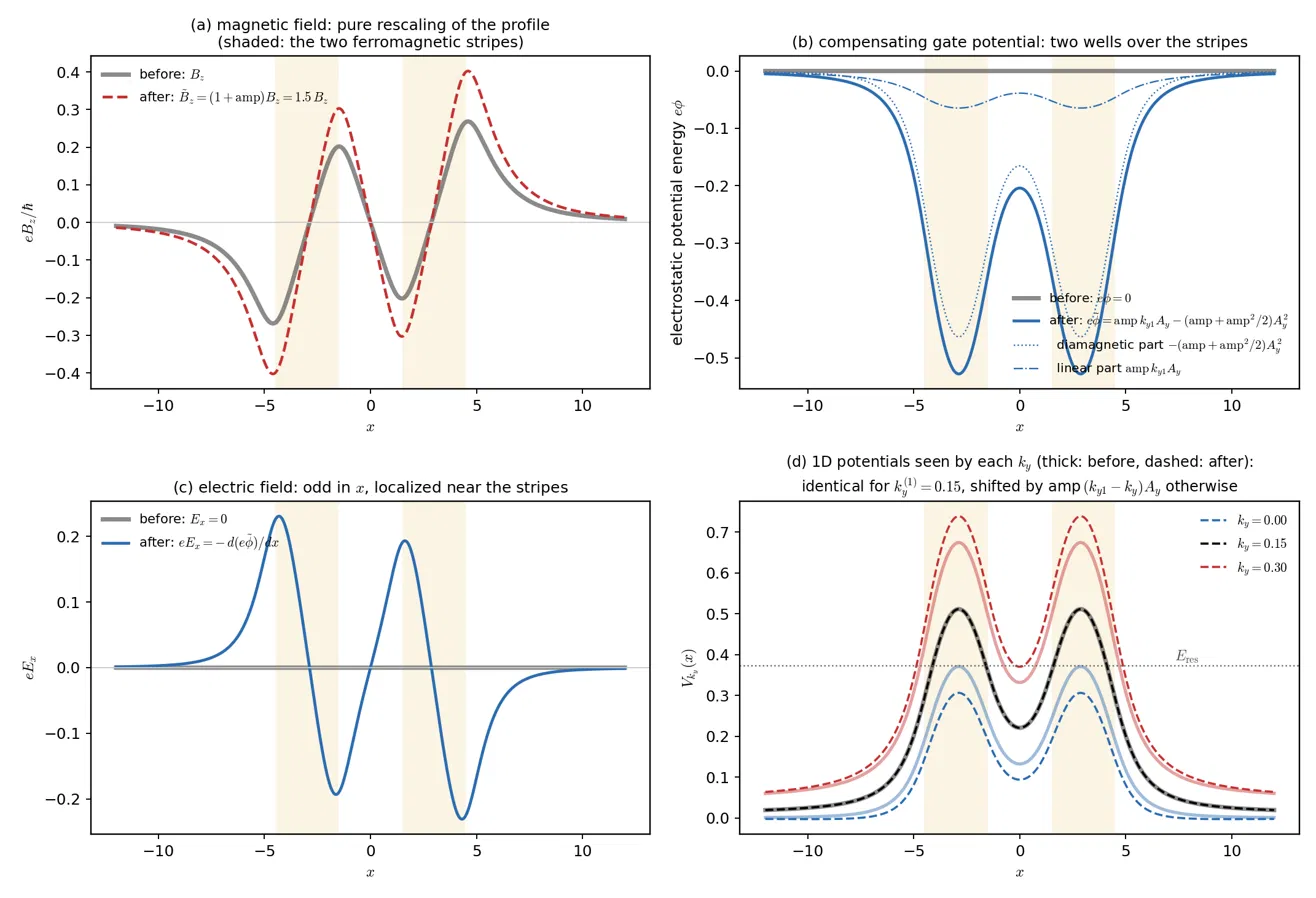}
    \caption{Fields for the double-stripe barrier and the deformation \eqref{eq:numfamA}--\eqref{eq:numfamphi} with $k_{y1}=0.15$, $a=0.5$. (a) Magnetic field $B_z$ before and after (the profile is rescaled by $1+a$; shaded: the two ferromagnetic stripes). (b) Compensating potential $e\tilde\phi$ with its linear-in-$A_y$ and diamagnetic ($A_y^2$) parts. (c) Electric field $eE_x$, odd in $x$ and localized near the stripes. (d) Effective 1D potentials $V_{k_y}$ for several $k_y$ (thick: before; dashed: after): identical for $k_y=k_{y1}$, shifted by $a(k_{y1}-k_y)A_y$ otherwise.}
    \label{fig:fields-res}
\end{figure*}

The deformation family is the one-parameter family
\begin{eqnarray}
\tilde A_y&=&(1+a)A_y,\label{eq:numfamA}\\
e\tilde\phi&=&a\,k_{y1}A_y-\Bigl(a+\frac{a^2}{2}\Bigr)A_y^2,\label{eq:numfamphi}
\end{eqnarray}
i.e.\ $\delta A_y=aA_y$ in Eq.~\eqref{eq:slabtrick}, with $a\equiv\mathrm{amp}$ the deformation amplitude (we write $\mathrm{amp}$ in the figures). It is localized (the leads are untouched), the magnetic field is simply rescaled, $\tilde B_z=(1+a)B_z$, and
\begin{equation}
\tilde V_{k_y}(x)=V_{k_y}(x)+a\,(k_{y1}-k_y)A_y(x),
\label{eq:numshift}
\end{equation}
which is Eq.~\eqref{eq:slabshift} for this $\delta A_y$. The value $a=-1$ switches the magnetic field off: $\tilde B_z=0$ and the barrier is purely electrostatic (an ``electrostatic twin''), which retains the protected channel only. The fields before and after the deformation are shown in Fig.~\ref{fig:fields-res}. For $k_{y1}=0.15$ and $a=0.5$ the maximal field grows from $|B_z|_{\max}=0.268$ to $0.402$, and the compensating potential $e\tilde\phi$ consists of two wells over the stripes (minimum $-0.528$ at $x=\mp2.88$), with $|eE_x|_{\max}=0.231$; $E_x$ is odd in $x$.

\subsubsection{Exact invariance of the protected channel}

The scattering problem for fixed $(E,k_y)$ was solved by Numerov integration of Eq.~\eqref{eq:slabradial} from $x=+L$ with an outgoing wave, backward to $x=-L$, giving the complex amplitudes $t,r$ and $T=(k_R/k_L)|t|^2$. For $k_y=k_{y1}$, $V_{k_{y1}}$ coincides before and after to $\sim4\times10^{-16}$, and the complex transmission amplitude $t(E)$ (not just $T$) is identical for all energies to $\sim10^{-13}$, for any $a$. Equivalently, any wave packet incident at fixed $k_{y1}$ evolves identically. For $k_y\neq k_{y1}$ the transmission changes (Figs.~\ref{fig:res1d}--\ref{fig:maps-res}).

A remark on saturation is needed when reading difference plots. A map of $\Delta T=\tilde T-T$ is also close to zero where $T\approx0$ or $T\approx1$ for any $k_y$, so a small $\Delta T$ alone is not a sign of protection. The protected line $k_y=k_{y1}$ has to be recognized by the vanishing of the \emph{relative} change of the suppressed quantity, $|\Delta\ln T|$ (or $|\Delta\ln R|$ where $T\approx1$), which vanishes only there (Fig.~\ref{fig:maps-res}).

\subsubsection{Resonance flow}

\begin{figure*}
    \centering
    \includegraphics[width=\textwidth]{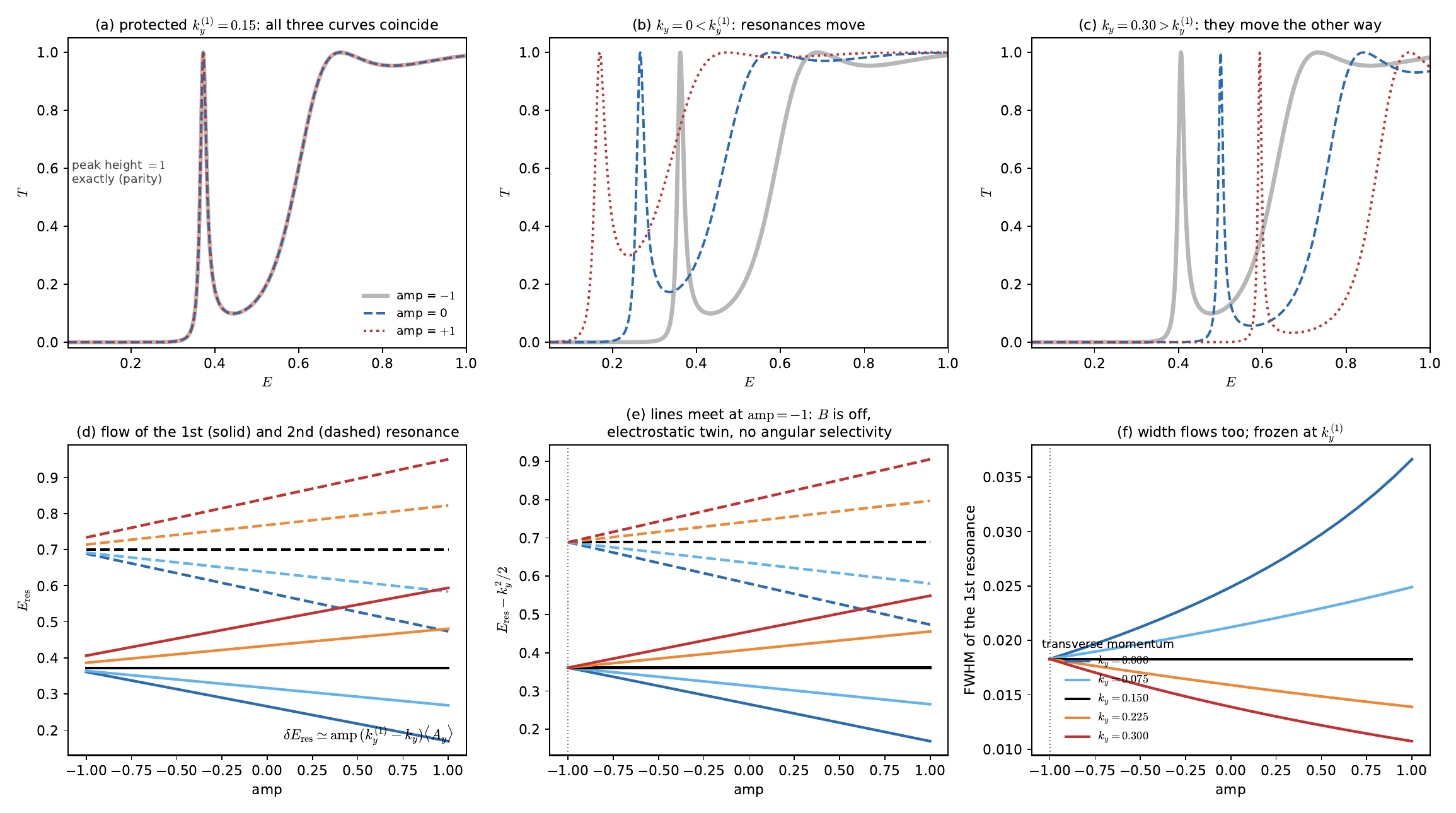}
    \caption{Resonance flow in the double-stripe barrier for $k_{y1}=0.15$. (a)--(c) $T(E)$ at $k_y=k_{y1}$, $0$ and $0.30$ for several amplitudes $a=\mathrm{amp}$: all curves coincide in (a). (d) Positions of the first (solid) and second (dashed) resonance versus $a$ for five $k_y$; the line for $k_y=k_{y1}$ is flat. (e) $E_{\rm res}-k_y^2/2$: all lines meet at $a=-1$ (magnetic field off, electrostatic twin). (f) Resonance width (FWHM) versus $a$; frozen at $k_{y1}$.}
    \label{fig:res1d}
\end{figure*}

Because the transmission at fixed $k_y$ is a one-dimensional scattering problem in the potential \eqref{eq:numshift}, the resonance positions $E_{\rm res}(k_y)$ respond to the deformation in a controlled way. To first order in $a$, perturbation theory for a quasi-bound state in the well gives
\begin{equation}
\delta E_{\rm res}\simeq a\,(k_{y1}-k_y)\,\langle A_y\rangle_{\rm res},
\label{eq:resflow}
\end{equation}
where $\langle A_y\rangle_{\rm res}$ is the average of $A_y$ over the resonant state. We find $\langle A_y\rangle\approx-0.63$ and $-0.72$ for the first and second resonances, close to the average over the cavity ($-0.62$ to $-0.64$). Fig.~\ref{fig:res1d} shows: (a) $T(E)$ at $k_y=k_{y1}$, where all curves for different $a$ coincide, with the first resonance at $E_{\rm res}=0.372295$ and $\Gamma_{\rm FWHM}=0.0183$ for $k_{y1}=0.15$, and a second resonance near $0.70$; (b,c) the same at $k_y=0$ and $k_y=0.30$, where the resonances move in opposite directions for $k_y<k_{y1}$ and $k_y>k_{y1}$ (the sign of the shift in Eq.~\eqref{eq:resflow} is that of $a\,(k_{y1}-k_y)\langle A_y\rangle$); (d) the resonance flow versus $a$ for five values of $k_y$, which is linear for small $|a|$ and has a fixed point at $k_y=k_{y1}$ (horizontal line); (e) $E_{\rm res}-k_y^2/2$ versus $a$: all lines meet at $a=-1$, where the magnetic field is off and the problem reduces to an electrostatic barrier, for which the potential is independent of $k_y$ apart from the trivial shift $k_y^2/2$; (f) the width flows too and is frozen at $k_{y1}$. The peak value of $T$ at resonance is exactly $1$ for any $k_y$ and $a$, because $V_{k_y}(x)$ is even in $x$ (the double-stripe potential and the compensation preserve the parity), so that the resonance is a perfect-transmission peak, and the effect of the deformation is a pure displacement and a change of the width of the peak, not of its height.

The point $a=-1$ provides a nontrivial numerical check. There $\tilde V_{k_y}=\tfrac12k_y^2-k_{y1}A_y+\tfrac12A_y^2$ depends on $k_y$ only through the additive constant $k_y^2/2$, so $\tilde T$ is a function of $E-k_y^2/2$ and $\partial_{k_y}\ln\tilde T=-k_y\,\partial_E\ln\tilde T$. At $k_y=k_{y1}$ we have $\tilde T=T$ for all $E$, hence $\partial_{k_y}\ln\tilde T|_{k_{y1}}=-k_{y1}\,\partial_E\ln T|_{k_{y1}}$; at a test energy the two sides are $+5.5361$ and $+5.5360$.

\subsubsection{Maps $T(E,k_y)$}

\begin{figure*}
    \centering
    \includegraphics[width=\textwidth]{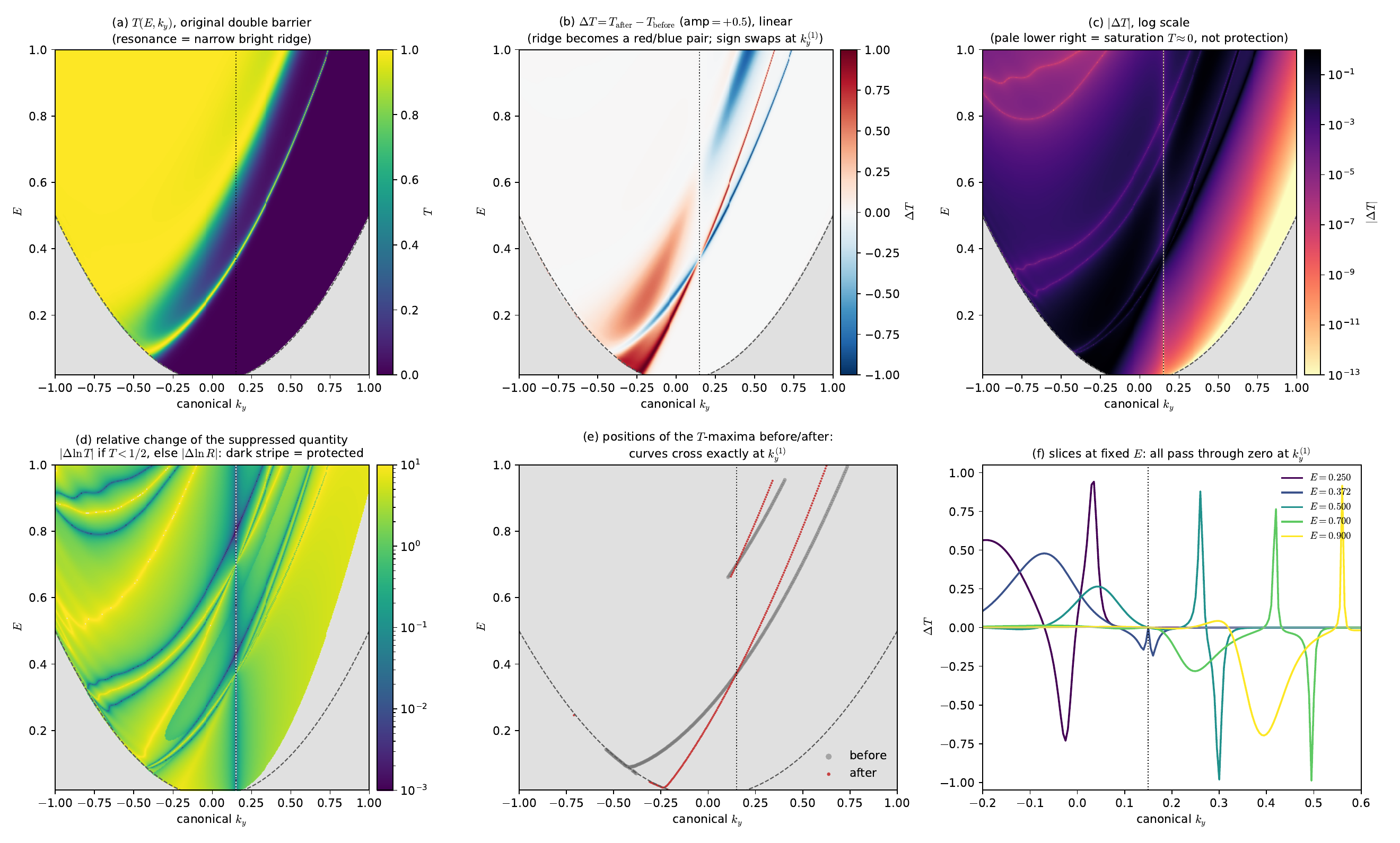}
    \caption{Transmission $T(E,k_y)$ of the resonant barrier ($k_{y1}=0.15$, $a=0.5$). (a) $T$ before deformation. (b) $\Delta T=\tilde T-T$, linear scale. (c) $|\Delta T|$, log scale. (d) Relative change $|\Delta\ln T|$ of the suppressed quantity, which isolates the protected line $k_y=k_{y1}$ from the saturated regions. (e) Positions of the resonance ridges before and after. (f) Slices $\Delta T(k_y)$ at fixed energy; all pass through zero at $k_{y1}$.}
    \label{fig:maps-res}
\end{figure*}

Fig.~\ref{fig:maps-res} shows the angle--energy dependence of the transmission: (a) $T(E,k_y)$ for $a=0$, where the resonances are narrow ridges whose position depends on $k_y$; (b) $\Delta T$ on a linear scale; (c) $|\Delta T|$ on a logarithmic scale; (d) the relative change $|\Delta\ln T|$ of the suppressed quantity; (e) the ridge positions before and after; (f) slices $\Delta T(k_y)$ at fixed $E$, all passing through zero at $k_{y1}$. The deformation moves the ridges sideways in the $(E,k_y)$ plane around the vertical line $k_y=k_{y1}$, which stays fixed, while the ridge of the unprotected channels moves through $\Delta T\sim1$ (the peak is shifted by more than its width). It is therefore a very sensitive probe: a change of the resonance position by less than the width $\Gamma$ gives a full change of $T$ at fixed energy.

\subsubsection{Resonant beam}

\begin{figure*}
    \centering
    \includegraphics[width=0.75\textwidth]{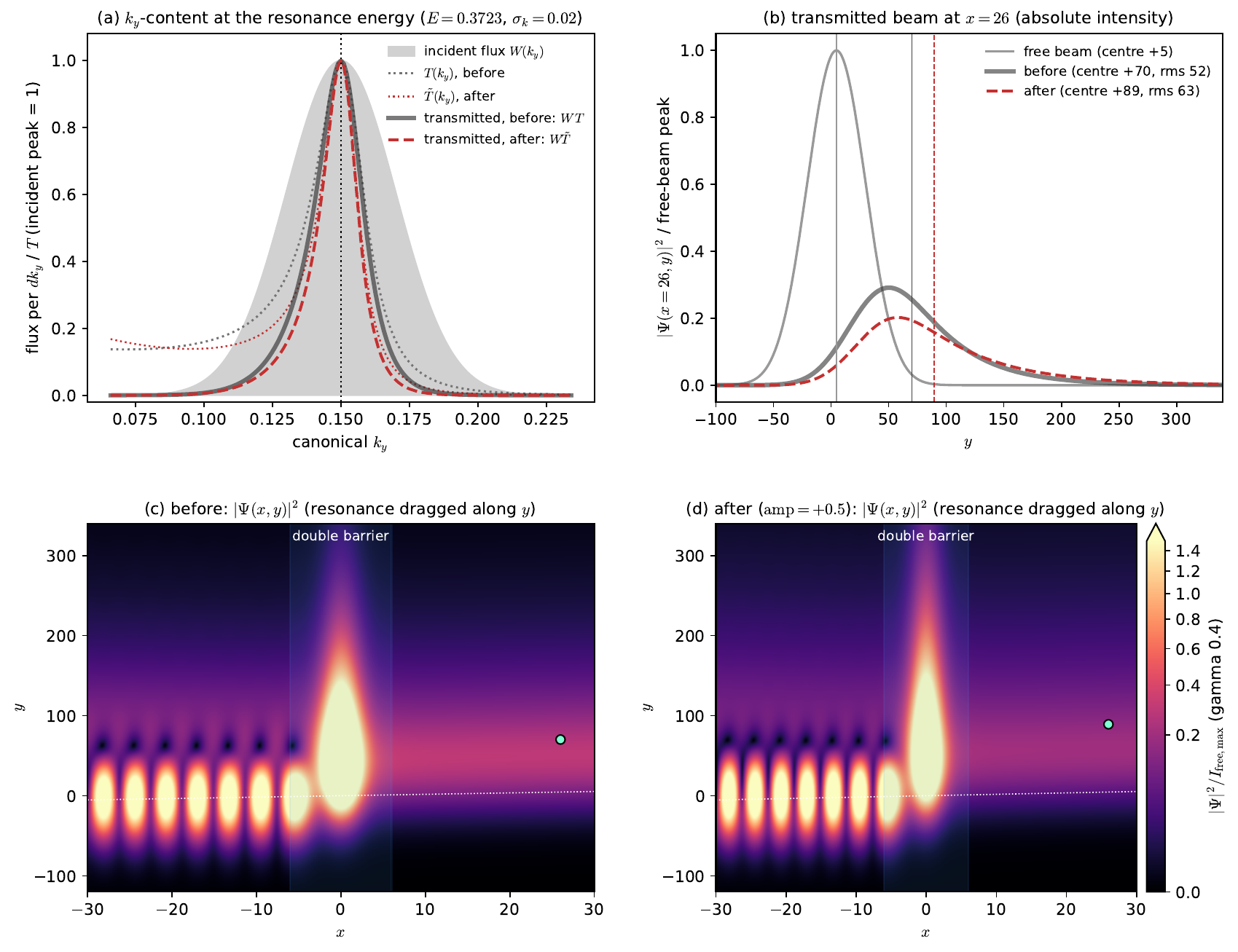}
    \caption{Beam at the resonance energy $E=E_{\rm res}$, $k_{y1}=0.15$, $\sigma_k=0.02$, $a=0.5$. (a) $k_y$-content of the beam at the resonance energy. (b) Transmitted profile at $x=26$ (absolute intensity). (c,d) $|\Psi(x,y)|^2$ before and after: the transmitted beam is dragged along $y$ and its transmitted fraction is reduced.}
    \label{fig:beam-res}
\end{figure*}

The $k_y$-dependence of the resonance has two observable consequences for a beam $\Psi(x,y)=\int g(k_y)\psi_{k_y}(x)e^{ik_yy}dk_y$ with a narrow Gaussian distribution $g$ around $k_{y1}$ at the resonance energy (we take $g=\sqrt{W/k_x}$ so that the incident flux density is $W(k_y)$). First, the resonance acts as a filter in angle: at $E=E_{\rm res}$ and $\sigma_k=0.02$ the transmitted fraction is $0.487$ before and $0.393$ after, because the width of the resonance in $k_y$ is comparable to $\sigma_k$ and the neighbouring $k_y$ components are detuned by Eq.~\eqref{eq:resflow}. The total transmitted fraction changes at first order in $a$ whenever $\sigma_k$ is not small compared to the width of the resonance in $k_y$: for $\sigma_k=0.002,\,0.004,\,0.008,\,0.012,\,0.02,\,0.04$ (at $a=0.5$) we find $\Delta T_{\rm tot}=-2.4,\,-6.7,\,-12.9,\,-16.2,\,-19.3,\,-19.9\,\%$. Protection is therefore a property of the single channel $k_{y1}$, and not of a beam of finite angular width. Second, the transmitted beam is displaced laterally: the centroid at $x=26$ (free propagation $+4.6$) moves from $+70.2$ before to $+89.2$ after, with an rms width increasing from $52$ to $63$ (Fig.~\ref{fig:beam-res}). The corresponding shift follows from $\partial_{k_y}\arg t$ (a Goos--H\"anchen-like shift) and for non-resonant propagation agrees with the classical drift $-a\int A_y/k_x\,dx$. At resonance the transverse drift $\partial_{k_y}E_{\rm res}$ is $0.783$ for $a=0$, $1.100$ for $a=0.5$ and $0.4665$ for $a=-0.5$, corresponding to $\partial_{k_y}\arg t=-95.2,\,-134.4,\,-55.9$, in agreement within $\sim10\%$ with the estimate $-2E'_{\rm res}/\Gamma=-85.7,\,-120.3,\,-51.0$. The ``resonance dragged along $y$'' is clearly seen in $|\Psi(x,y)|^2$.

For comparison, for a single (non-resonant) stripe, a beam incident at $E=0.4$ with $k_{y1}=0.15$, $\sigma_k=0.1$ and $a=0.5$ has its mean outgoing angle changed from $6.4^\circ$ to $5.75^\circ$, centroid shift from $+10.8$ to $+13.9$ and transmitted fraction from $0.4707$ to $0.4776$: these effects are small compared with the resonant case, where the sharp $k_y$-dependence of the resonance position amplifies the sensitivity.

\subsubsection{Gate realization and tolerances}

\begin{figure*}
    \centering
    \includegraphics[width=0.9\textwidth]{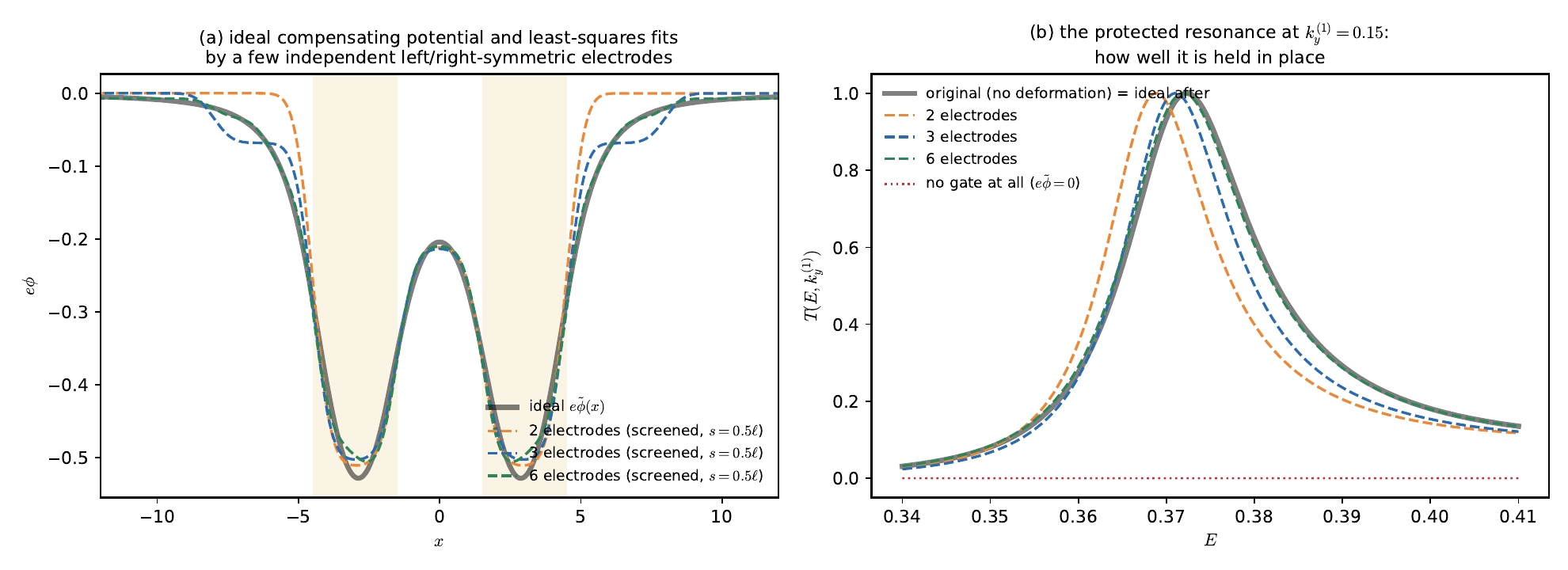}
    \caption{Gate realization. (a) Ideal compensating potential $e\tilde\phi(x)$ and least-squares fits by $2$, $3$ and $6$ independent left--right symmetric electrodes (Gaussian screening $s=0.5\ell$). (b) The protected resonance at $k_{y1}=0.15$ for each realization, showing how well it is held in place.}
    \label{fig:gates}
\end{figure*}

The compensating potential \eqref{eq:numfamphi} is not produced by the ferromagnetic stripes alone: it needs electrostatic gates, which also must cover the region \emph{between} the stripes. We tested what accuracy is required by fitting $e\tilde\phi(x)$ for $k_{y1}=0.15$, $a=0.5$ with a few independent, left--right symmetric Gaussian-screened electrodes of width $s=0.5\ell$ (Fig.~\ref{fig:gates}). With $2,\,3,\,6$ electrodes the resonance of the protected channel is shifted by $-0.19,\,-0.07,\,-0.02\,\Gamma$ respectively. If the amplitude of the ideal potential is off by $1\%$ ($5\%$) the resonance shifts by $\mp0.17\Gamma$ ($\mp0.84\Gamma$), and smoothing of the potential by the screening length $s=0.25,\,0.5,\,1.0\,\ell$ gives $-0.10,\,-0.38,\,-1.4\,\Gamma$. If only the regions over the stripes are gated the shift is $+7.8\,\Gamma$, i.e.\ the protection is lost entirely. For a resonance of width $\Gamma\simeq0.018$ this means that the gate amplitude must be controlled at the level of a few percent, or better if the resonance is to be held within a fraction of its width. We suggest, but have not tested, a calibration procedure: tune the gate voltages until the transmission at $(E,k_{y1})$ coincides with that of the original device, and then use the other channels as a diagnostic.

\subsubsection{Limitations}

These results hold for the infinite strip in $y$, with a $k_y$-conserving, ballistic and non-interacting electron; scattering, finite temperature, and the lateral confinement of the strip (hard walls break translation invariance in $y$ and couple the transverse modes) were not included. The exact protection applies to a single $k_{y1}$, and for beams or finite-temperature distributions the effect on the total transmission is generically of first order in $a$.

% The \nocite command causes all entries in a bibliography to be printed out
% whether or not they are actually referenced in the text. This is appropriate
% for the sample file to show the different styles of references, but authors
% most likely will not want to use it.
%\nocite{*}

\bibliography{apssamp}% Produces the bibliography via BibTeX.

\end{document}